\documentclass[12pt,english]{article}
\usepackage{amssymb,amsmath}
\usepackage[spanish,activeacute]{babel}
\usepackage[latin1]{inputenc}

\usepackage{array}
\usepackage{float}
 
\AtBeginDocument{}

\def\'#1{{\accent19\ifx #1i \i\else #1\fi}}

\def\be{\begin{equation}}
\def\ee{\end{equation}}
\def\bea{\begin{eqnarray}}
\def\eea{\end{eqnarray}}

\newcommand{\bolditQ}{\mbox{\it\boldmath$Q$\unboldmath}}

\newcommand{\boldmathPsi}{\mbox{\boldmath$\Psi$\unboldmath}}

\newcommand{\boldmathphi}{\mbox{\boldmath$\phi$\unboldmath}}

\newbox\Ancha
\catcode`@=11
\newdimen\ex@
\title{ Representation of quantum field theory  in an extended spin space and fermion mass hierarchy}

\author{J. Besprosvany and R. Romero}

\date{Instituto de F\'{\i}sica, Universidad Nacional Aut\'onoma de M\'exico,
Apartado Postal 20-364, M\'exico 01000, D. F., M\'exico }

\begin{document}

\renewcommand{\tablename}{Table}
\renewcommand{\abstractname}{Abstract}
\renewcommand{\refname}{References}


\maketitle







\jot = 1.5ex
\def\baselinestretch{1.9}
\parskip 5pt plus 1pt

\begin{abstract}

We consider a matrix space based on the spin degree of freedom,    describing both a Hilbert state space,  and  its corresponding symmetry  operators. Under the requirement that the Lorentz symmetry be kept,
at given dimension,    scalar symmetries, and their representations are determined. Symmetries are flavor or  gauge-like,  with fixed chirality.
After    spin  0, 1/2, and 1  fields  are obtained    in this    space, we construct  associated  interactive gauge-invariant  renormalizable terms,  showing  their  equivalence to a Lagrangian formulation, using as example the previously studied (5+1)-dimensional case, with many standard-model connections.
At 7+1 dimensions, a  pair of Higgs-like  scalar Lagrangian is obtained naturally producing    mass  hierarchy within a fermion flavor  doublet.

\end{abstract}
\vskip 2cm \centerline{PACS: 12.60.-i, 11.15.-q, 12.10.Dm,
11.30.Rd .}


\baselineskip 22pt\vfil\eject \noindent

  \section {Introduction}

The current theory of elementary particles, the standard model
(SM), is successful in describing their behavior, but it is
phenomenological. The origin  of  the interaction groups, the
particles' spectrum and representations, and parameters has
remained largely unexplained. A unified theory can aim to  build physical objects from the most
elementary ones.  The generalization of
features of the model into larger structures with a unifying
principle has suggested connections among the observables. Thus,
additional spatial dimensions in Kaluza-Klein theories are
associated with gauge symmetries, and
 larger gauge groups in grand-unified theories (GUTs)$\cite{unification}$ put some restrictions
on them.

Particles and interactions obey Lorentz-scalar and  local symmetries,  associated to gauge groups. 
    The  fundamental representation of the Lorentz group  is physically manifested in elementary-particle fermions,
  while    the vector representation corresponds to interaction bosons.
On the other hand, fermions  occupy
 the scalar-group fundamental representation, and vector particles the adjoint.   In addition, the description and quantification\footnote{Expressed in quantum numbers within quantum mechanics.} of
 particles and interactions  have similar consistency requirements,    as restrictions on the representations
from unitarity. These notable similarities and connections  between the existing particles' discrete degrees of freedom   point to
   a common origin, and hence, a simple composite description.

  Indeed, a  shared  vector  space,  was    proposed\cite{Jaime,JaimeB} that generalizes spin, and  accommodates scalar and  Lorentz
degrees of freedom;  at given dimension[d], this  space constrains  the symmetries and representations, and  its generators  in the dimensions beyond 3+1 are associated with scalar
symmetries.
 While only a simple  admixture of Lorentz and scalar groups is permitted  by the Coleman-Mandula theorem\cite{Coleman}, additional non-trivial  information is obtained from the spin-space   scheme, as chiral and vector characterizations emerge from the    symmetries  and particle representations.
   Similarly to the  supersymmetry case\cite{wess},   the dimension of the space  constrains the
   particle spectrum;  as we will explain, the interactions are also constrained.

Within a Kaluza-Klein framework, this extension may be  viewed as a consequence of the spatial components being  frozen. Conceptually, the  matrix construction  stems from incremental direct products with $2\times 2$ matrices,  suggesting   the  discrete Hilbert space used is built up from the most elementary degrees of freedom (e. g., q-bits or spin-1/2 particles.)

Although standard SM extensions provide additional information on it, many puzzles remain unsolved. With its bottom-up approach, this model reduces the available groups and representations to fit particles and their quantum numbers,  in contrast, e. g., to  the  representation choices available in GUTs, and to the multiplicity of  compactification options that plagues   strings.


  While this scheme was used before to derive information on coupling constants\cite{bespro9m1}, SM representations \cite{bespro9m1,Besprosub}, and relations between electroweak  boson masses\cite{Besprosub}, a formal treatment  to produce    an interactive model was missing. In this paper, we construct  step-by-step gauge- and Lorentz-invariant terms  from    fields  within representations and symmetries that derive  from the extended spin space, which   translates into a Poincar\'{e}-invariant Lagrangian theory.
In particular,
 we  show  formally the equivalence of a gauge-invariant  field theory, written in
such a space, and a standard formulation, thus extending and complementing previous work\cite{bespro9m1,Besprosub,BesproMosh};   each vertex type exhibits  particular features.  We also find that the scalar fermion-scalar term in (7+1)-d implies a hierarchy in the fermion masses.

The paper is organized as follows.
 In Section 2, we review   the   spin-space extension, based on the Dirac equation, and its connection
  to   a matrix space.
In particular,   we  present  its   elements' classification, using a Clifford algebra, under the demand that Lorentz symmetry be
maintained;  in such a space,  spinors belong to the scalar-group fundamental  representation,  while  vectors  to the adjoint representation\cite{Jaime,JaimeB,bespro9m1,Besprosub,BesproMosh}. In Section 3,   generalized  fields  and symmetries are  expressed in this space (using as example  the   (5+1)-d  case).
 In Section 4,  these are used to construct  a  gauge-invariant interactive theory,  showing that it can be formulated in terms of a standard  Lagrangian; we deal with vector-scalar, fermion-vector and vector-vector vertices,
  using the obtained groups  $\rm SU(2)_L \bigotimes U(1)_Y$ in  (5+1)-d, and correct chirality.
  In Section 5,       fermion-scalar vertices are obtained   in   7+1 d.   Higgs-like scalars emerge that lead, through the Higgs mechanism, to fermion masses in a flavor doublet;   Yukawa couplings naturally generate a fermion-mass hierarchy. In Section 6, we summarize relevant points in the paper.







  Other   investigations similarly rely on
 the spin  degree of freedom in SM  extensions\cite{spin,Chisholm},  in trying to understand its still unresolved questions.
These use an  algebraic spinor
represented by a matrix, where the common feature of this type
of model building is the use of the structure within an associated Clifford-algebra
space. In  four dimensions,
   a $4\times4$ matrix
connects  to the $(3+1)$-d Clifford algebra ${\mathcal C}_{4}$.
 Each column in the matrix is a left ideal of the algebra. This allows for   operators acting from the right, and
such transformations are usually associated with gauge groups.
   To take
account of  the SM particle multiplets and gauge groups, one
introduces extra spacetime dimensions. Different choices are made for the nature of the left ideals, the spacetime
dimensions, and
symmetry transformations,
which leads to different models with various degrees of applicability and
phenomenological implications.
In Refs. \cite{Bracic:2005ic,Bregar2008}
models   based on Clifford objects in $13+1$  d  purport  to explain the origin of quark and lepton families.
In Ref.  \cite{Trayling:2001kd} an algebraic spinor
of ${\mathcal C}_{7}$ is used to represent one family of quarks and leptons,
with Poincar\'e and gauge transformations restricted to act from the
left and right, respectively.

Other types of   models  include gravity and   are geometric
in nature. Thus, the fundamental Clifford algebra relation,
usually taken as a real algebra, is given in terms of an abstract vector
basis $\{e_{\mu}\}$ as
\begin{eqnarray}
\label {geometrical}
e_{\mu}e_{\nu}+e_{\nu}e_{\mu}=g_{\mu\nu}  ,
\end{eqnarray} without reference to the gamma matrices.
To cite some recent examples (in no manner an exhaustive list), in
Refs.  \cite{Lu2011} and \cite{Nesti2009} models include the
SM gauge groups and gravity, the former based on ${\mathcal C}_{6}$,
and  the latter  on ${\mathcal C}_{3+1}$,  which assumes a column spinor   within an algebraic
matrix. Ref. \cite{Pavsic2010} also advances a model including
gauge and gravity fields, motivated by strings and branes models, and
set up in a $16$-d Clifford space.


\section{Gamma-matrix symmetry classification}
  The Dirac equation  formulated over the  matrix $\Psi$ (and corresponding conjugate equation)
\begin{eqnarray}
\label {JaimeqDi} \gamma_0( i \partial_\mu\gamma^\mu -M)\Psi ={ 0},
\end{eqnarray}
may be used as framework for the classification of states and operators in an extended space,\footnote{We assume throughout $\hbar=c=1$, and 4-d diagonal  metric elements $g_{\mu\nu}=(1,-1,-1,-1).$}    and  study symmetry transformations.
 It    also  generates free-particle  fermion and bosons  on   the extended space.

These matrices  generate an
  algebra, and may   be also viewed in terms of their bra-ket components:
\begin{eqnarray}
\label {Jaimeqnext}  \Gamma= \sum c_{a b}|a \rangle \langle  b |,
\end{eqnarray}
with $c_{a b}$   $c$-numbers.
The  dot product between the elements $\Gamma_a$, $\Gamma\textbf{}_b$ can be defined using the trace
\begin{eqnarray}
\label {dotprod} {\rm  tr \ } \Gamma_a^\dagger  \Gamma_b.
\end{eqnarray}Assuming $\Gamma$ in Eq. \ref{Jaimeqnext}
  to be unitary we obtain the condition\cite{Jaime,JaimeB}  on  the $c_{a b}$  values
 \begin{eqnarray}
\label {unitarity}  \Gamma^\dagger\Gamma= \sum c^*_{b a} c_{b c}  = \delta_{ac}.
\end{eqnarray}

 Implicit in the $| a\rangle \langle b |$ matrix construction is the appropriate  transformation  operators $U$  acting on field states $\Psi$; these  can  generically  be characterized by the expression
\begin{eqnarray}
\label  {transfoGen}
 \Psi\rightarrow U \Psi U^\dagger.
\end{eqnarray} We  show next that a matrix $\Gamma$  can be associated to either $\Psi$ and
  $U$, the latter representing both Lorentz and scalar  symmetries.
We also show that a  4-dimensional Clifford  matrix  subalgebra is obtained,
  implying  spinor up to  bi-spinor elements, thus
    vectors and scalar fields, can be described.

  An operator $Op$ within this space  characterizes a state  $\Psi$ with the eigenvalue rule
\begin{eqnarray}
\label {OpAct}
[Op,\Psi]=\lambda  \Psi,
 \end{eqnarray}
 consistent with the hole interpretation, and anticipating a second-quantization description. For example, an on-shell boson may be constructed by two fermion components with   positive   frequencies   $\psi_1(x)$, $\bar\psi_2 (x)$  through $\psi_1(x)\bar\psi_2(x)$, following Eq. \ref{Jaimeqnext}, with $\bar\psi_2 (x)$ describing an antiparticle.


  If Eq. \ref{JaimeqDi},  keeping  $\mu=0,...,3,$  is assumed within  the larger Clifford
algebra\footnote{Understood here also as a matrix space. } ${\mathcal C}_{N},$
 $\{ \gamma_\eta ,\gamma_\sigma \} =2 g_{\eta\sigma}$, $\eta,\sigma=0,...,N-1$, with $N$ the
 (assumed even) dimension,
 whose structure is
  helpful in classifying the available symmetries $U$,  and
  solutions $\Psi$, both
represented  by  $2^{N/2}\times 2^{N/2}$
 matrices.
 The 4-d Lorentz symmetry is maintained, and  uses the generators
\begin{eqnarray} \label {sigmamunu}
 \sigma_{\mu\nu}=\frac{i}{2}[\gamma_\mu ,\gamma_\nu ],
 \end{eqnarray}
   where $\mu ,\nu =0,...,3.$
  $U$  contains also $ \gamma_a$, $a=4,...,N-1$,  and their
 products as possible symmetry
 generators.
   The $N=4$ case was analyzed in Ref. \cite{Jaime,JaimeB}, $N=6$ in    $\cite{Jaime}$, $\cite{JaimeB}$, \cite{Besprosub}, and $N=10$ in \cite{bespro9m1}.
Indeed, the  latter elements are  scalars for they commute
 with the Poincar\'e generators, which contain  $\sigma_{\mu\nu}$, and
they are also  symmetry operators  of  the massless  Eq.
\ref{JaimeqDi},  bilinear in  $\gamma_\mu,$ $\mu=0,...,3$
 which is not
  necessarily the case for mass terms (containing  $\gamma_0$).
 In addition, their products with
   $\tilde\gamma_5=-i\gamma_0\gamma_1\gamma_2\gamma_3$  are Lorentz  pseudoscalars.
 As
 $[\tilde\gamma_5,\gamma_a]=0$,
  we can classify the (unitary) symmetry algebra as
  ${\mathcal S}_{N-4}={\mathcal S}_{(N-4)R} \bigotimes {\mathcal S}_{(N-4)L},$
  consisting of the projected
 right-handed ${\mathcal S}_{(N-4)R}=
\frac{1}{2}(1+\tilde\gamma_5){\rm U}(2^{(N-4)/2})$ and left-handed ${\mathcal
S}_{(N-4)L}= \frac{1}{2}(1-\tilde\gamma_5)U(2^{(N-4)/2})$
  components, where ${\rm U}(M)$ is  a representation
  of the $M$-unitary group in ${\mathcal C}_{N}$. Its reduced form $\tilde {\rm U}(2^{(N-4)/2})\subset
    {\mathcal C}_{N-4}$, with ${\mathcal C}_{N} = {\mathcal C}_{4}\bigotimes {\mathcal C}_{N-4},$ is
    the irreducible fundamental representation. The operator algebra was
 described in Refs. \cite{bespro9m1} and \cite{BesproMosh}.


A state $\Psi$ is  classified in accordance with the above symmetry generators that emerge from the Clifford algebra.    For given  dimension $N$, any matrix element  representing a state is obtained by combinations
of products of one or two $\gamma_\mu$, and elements of ${\mathcal
S}_{N-4}$, which define, respectively, their Lorentz (as for 4-d)
and scalar-group representation.
There  is a finite number of partitions on the matrix space for the states and symmetry operators, consistent with Lorentz symmetry.
 The variations of  the symmetry algebra
are defined  by   the projection operators ${\mathcal P}_P, \ {\mathcal
P}_S\in {\mathcal S}_{N-4}$
with $[{\mathcal P}_P,{\mathcal P}_S]=0$; ${\mathcal P}_P$ acts on
  the Lorentz generator
  \begin{eqnarray}
\label{Lorentgen} {\mathcal
P}_P[\frac{1}{2}\sigma_{\mu\nu}+i(x_\mu\partial_\nu-
x_\nu\partial_\mu)],
\end{eqnarray}and ${\mathcal P}_S$ on the symmetry operator space
  \begin{eqnarray}
\label{Sprima}{\mathcal S}_{N-4}^\prime={\mathcal P}_S {\mathcal S}_{N-4},\end{eqnarray}  leading to    projected  scalar generators $I_a={\mathcal P}_S I_a$, so that  they  determine, respectively, the
Poincar\'e generators and  the scalar groups.



\def\baselinestretch{1.1}
\renewcommand{\arraystretch}{.85}

\begin{table}[h]
\center{ \begin{tabular}{||c|c||c ||} \hline\hline
 \raisebox{-1.0ex}{(*)} &    & \raisebox{-1.0ex}{   } \\  [0.2 cm]
\hline
 &     &  \\
   &    &  \\
   &    &  \\
 \raisebox{-0.2ex}{ }  & \raisebox{-0.2ex}{${\mathcal S}_{(N-4)R}\bigotimes {\mathcal C}_{4} $}  & \raisebox{-0.2ex} {}    \\ 
 &    &  \\
 &    &  \\
   &  \ \ \ \ \ \ \ \  \ \ \  { \small {{  }}}  & \ \ \ \ \ \ \ \ \ \ \ \ \ \  \ \ \ { \small {{ }}} \\
\hline\hline
 &    &  \\
 &    &  \\
 &    &  \\
  \raisebox{-1.5ex}{ }  & \ \ \ \ \  \raisebox{-1.5ex}{ }  \ \ \ \ \ &       \raisebox{-1.5ex}{ ${\mathcal S}_{(N-4)L}\bigotimes {\mathcal C}_{4} $} \\
  &    &  \\ 
  &    &  \\
   &    &  \\
   &  \ \ \ \ \ \ \ \ \ \ \ \  {{ } }  & \ \ \ \ \ \ \ \ \ \ \ \ \ \ \  \ \ \ { \small {{   }}} \\
\hline\hline
\end{tabular}}\\[.5cm] {{1(a)} }
\label{tab:tablejbFBVMatr}
\end{table}
\renewcommand{\arraystretch}{1}
\normalsize

 \def\baselinestretch{2.1}
\normalsize

\def\baselinestretch{1.1}
\renewcommand{\arraystretch}{.85}

\begin{table}[h]
\center{ \begin{tabular}{||c|c||c ||} \hline\hline
 \raisebox{-1.0ex}{  (*)}  & \raisebox{-1.0ex}{ F } & \raisebox{-1.0ex}{ F } \\  [0.2 cm]
\hline
&    &  \\
   &    &  \\
   &    &  \\
 \raisebox{-0.2ex}{F}  & \raisebox{-0.2ex}{V} & \raisebox{-0.2ex}{S,A}    \\ 
 &    &  \\
 &    &  \\
   &  \ \ \ \ \ \ \ \  \ \ \  { \small {{  }}}  & \ \ \ \ \ \ \ \ \ \ \ \ \ \  \ \ \ { \small {{ }}} \\
\hline\hline
 &    &  \\
 &    &  \\
 &    &  \\
  \raisebox{-1.5ex}{F}  & \ \ \ \ \  \raisebox{-1.5ex}{S,A}  \ \ \ \ \ &   \ \ \ \ \ \  \ \ \  \raisebox{-1.5ex}{V } \ \ \ \ \ \  \ \ \ \ \  \\
  &    &  \\ 
  &    &  \\
   &    &  \\
   &  \ \ \ \ \ \ \ \ \ \ \ \  {{ } }  & \ \ \ \ \ \ \ \ \ \ \ \ \ \ \  \ \ \ { \small {{  }}} \\
\hline\hline
\end{tabular}}\\[.5cm] {{1(b)} } \\
\label{tab:tablejbFBVSOl}
\caption{ (a) shows the arrangement of symmetry operators $U$ in matrix space of arbitrary dimension $N$, after projection over  ${\mathcal S}_P$, with left-handed and right-handed operators subspaces\cite{bespro9m1}; (*) represents the matrix subspace containing  the projector $1-  {\mathcal P}_{S}=1-{\mathcal P}_{P}$; its choice within the right-handed symmetry components is arbitrary.
 (b) shows  the arrangement of matrix  solutions $\Psi$  in  the  extended-spin model
is divided into four $\frac{N}{2}\times \frac{N}{2}$ matrix
 blocks,  containing  fermion (F), vector (and axial-)
(V), and scalar (and pseudo-), and antisymmetric  (S,A) terms. }
\end{table}
\renewcommand{\arraystretch}{1}
\normalsize
 \def\baselinestretch{2.1}
\normalsize
\baselineskip 22pt\vfil\eject \noindent

 The application of these operators follows  the operator rule in Eq. \ref{OpAct}, which assigns states to particular Lorentz and scalar group representations. For simplicity, we assume ${\mathcal P}_P ={\mathcal P}_S\neq 1$, as other possibilities are less plausible\cite{bespro9m1}.   Thus,  the Lorentz or scalar operators  act trivially on one side  of
solutions   of the form $\Psi={\mathcal P }_P\Psi ( 1-{\mathcal P }_P) $,  since
$(1-{\mathcal P }_P){\mathcal P }_P=0$, leading to  spin-1/2 states or states  belonging to the fundamental representation of the non-Abelian symmetry groups, respectively.

On  Table 1(a), we show schematically the organization of the symmetry operators, producing corresponding Lorentz and scalar  generators.
Table  1{(b)} also depicts  the resulting solution representations,
distributed according to their Lorentz classification: fermion, scalar, vector, and antisymmetric tensor.
 The matrices are classified  according to the   chiral  projection operators $\frac{1}{2}(1\pm\tilde \gamma_5)$, leading to
$N/2  \times   N/2$  matrix blocks   in ${\mathcal
C}_{N}$.  The space projected by  ${\mathcal P}_P ={\mathcal P}_S\neq 1$ is also depicted.

The chiral property of the fermion representations contrasts with the difficulty to reproduce it in traditional Kaluza-Klein extensions\cite{Witten}. In addition, when deriving a unitary subgroup SU$(M-4)$, for arbitrary $M$, departing from an extended Lorentz group  requires O$(2M-5,1)\supset$ SU$(M-4)\bigotimes$ O(3,1), while in our scheme, the subgroup  chain  can be chosen as    U$(M-4)\supset $ O(3,1)  $\bigotimes$ SU$(M-4)_{R} \bigotimes $ SU$(M-4)_{L}$.  This means lower dimensional spaces are sufficient to reproduce the SM groups,   reducing the representation sizes, and eliminating spurious degrees of freedom; in addition, the right- and left-handed group separation is possible for all dimensions.

While a grand-unified group limits the representations among which one must choose to put particles, in our case, the representations are determined.  Indeed, the specific combinations also emerge, corresponding to spin-1/2-fundamental and
vector-adjoint,  Lorentz and scalar groups representations, respectively; graphically,  vectors and scalar group elements occupy the same
matrix spots  (and similarly for fermions,) as seen in Tables 1(a) and 1(b).

%
%
%
%
In the  next Section,  we generalize these fields.


\section {Fields and symmetries in matrix space}

To construct interactive fields,  we start with free fields   within the   $(5+1)$-d space\cite{Besprosub}
 case  as   example, for  which we   highlight  predicted physical features.
There,  among  few choices,
${\mathcal P}_P=L$, with
$L=\dfrac{3}{4}-\dfrac{i}{4} (1+\tilde{\gamma}_{5} )\gamma^{5}\gamma^{6}-\dfrac{1}{4}\tilde{\gamma}_{5}$
is associated to the lepton number, and the resulting symmetry generators and particle spectrum fits
 the SM electroweak sector. Specifically, the projected symmetry space  also includes the  SU(2)$_L\bigotimes \rm U(1) _Y$  groups, with respective generators $I_i$ and hypercharge $Y$
\begin{eqnarray}
\label  {geneSU2Y}
I_{1}&=&\dfrac{i}{4} (1-\tilde{\gamma}_{5} )\gamma^{5}   \nonumber  \\
I_{2}&=&-\dfrac{i}{4} (1-\tilde{\gamma}_{5} )\gamma^{6}  \nonumber  \\
I_{3}&=&-\dfrac{i}{4} (1-\tilde{\gamma}_{5} )\gamma^{5}\gamma^{6}  \nonumber  \\
Y&=&-1+\dfrac{i}{2} (1+\tilde{\gamma}_{5} )\gamma^{5}\gamma^{6}.
\end{eqnarray}
We note that the $\rm SU(2)$ generators correctly contain  the projection operator $\dfrac{1}{2} (1-\tilde{\gamma}_{5})$, confirming the interaction's chiral nature, which also leads to chiral representations, a feature that results from
  nature of the matrix space under projector $L$ and the Lorentz group.  Under Eq. \ref{OpAct}, the action of these operators   on choices of free-particle states $\Psi$ is given on Table 2 together
with their quantum numbers.


 The question  on   what fixes  this extension's   dimension to derive SM groups and representations similarly applies to GUTs, as there is also an infinite number of possible groups that contain the SM.   The answers for both extensions hinge on  that the lowest dimension numbers already give relevant information, and on predictability  as, in our case, features as representations and chiral SU(2) are derived.

\def\baselinestretch{2.1}
\normalsize



\def\baselinestretch{1.1}
\renewcommand{\arraystretch}{.85}

\begin{table}[h]
\noindent \begin{raggedright}
\begin{tabular}{|>{\raggedright}p{0.13\textwidth}|>{\centering}m{0.44\textwidth}|>{\centering}p{0.07\textwidth}|>{\centering}p{0.05\textwidth}|>{\centering}p{0.05\textwidth}|>{\centering}p{0.05\textwidth}|>{\centering}p{0.08\textwidth}|>{\centering}p{0.05\textwidth}|}
\hline
Electroweak multiplets & States $\Psi$ & $I_{3}$ & $Y$ & $Q$ & $L$ & $\frac{i}{2}L\gamma^{1}\gamma^{2}$ & $L\tilde{\gamma}_{5}$\tabularnewline
\hline
Fermion doublet & $\begin{array}{c}
\frac{1}{8} (1-\tilde{\gamma}_{5})   (\gamma^{0}+\gamma^{3} )(\gamma^{5}-i\gamma^{6})\\
\frac{1}{8} (1-\tilde{\gamma}_{5})(\gamma^{0}+\gamma^{3} )(1+i\gamma^{5}\gamma^{6} )
\end{array}$ & $\begin{array}{r}
1/2\\
-1/2
\end{array}$ & $\begin{array}{r}
-1\\
-1
\end{array}$ & $\begin{array}{r}
0\\
-1
\end{array}$ & $\begin{array}{r}
1\\
1
\end{array}$ & $\begin{array}{r}
1/2\\
1/2
\end{array}$ & $\begin{array}{r}
-1\\
-1
\end{array}$\tabularnewline
\hline
Fermion singlet & $\frac{1}{8}$ $(1+\tilde{\gamma}_{5}  )
\gamma^{0}( \gamma^{0}+\gamma^{3}  ) ( \gamma^{5}-i\gamma^{6})$ & $0$ & $-2$ & $-1$ & $1$ & $1/2$ & $1$\tabularnewline
\hline
Scalar doublet & $\begin{array}{c}
\frac{1}{4\sqrt{2}}(1-\tilde{\gamma}_{5} )\gamma^{0}
(1-i\gamma^{5}\gamma^{6})\\
\frac{1}{4\sqrt{2}} (1-\tilde{\gamma}_{5} )\gamma^{0}
 ( \gamma^{5}+i\gamma^{6}  )
\end{array}$ & $\begin{array}{r}
1/2\\
-1/2
\end{array}$ & $\begin{array}{r}
1\\
1
\end{array}$ & $\begin{array}{r}
1\\
0
\end{array}$ & $\begin{array}{r}
0\\
0
\end{array}$ & $\begin{array}{r}
0\\
0
\end{array}$ & $\begin{array}{r}
-2\\
-2
\end{array}$\tabularnewline
\hline
Vector triplet & $\begin{array}{c}
\frac{1}{4}(1-\tilde{\gamma}_{5})\gamma^{0}(\gamma^{1}+i\gamma^{2})(\gamma^{5}-i\gamma^{6})\\
\frac{1}{2\sqrt{2}}(1-\tilde{\gamma}_{5})\gamma^{0}(\gamma^{1}+i\gamma^{2})\gamma^{5}\gamma^{6}\\
\frac{1}{4}(1-\tilde{\gamma}_{5})\gamma^{0}(\gamma^{1}+i\gamma^{2})(\gamma^{5}+i\gamma^{6})
\end{array}$ & $\begin{array}{r}
1\\
0\\
-1
\end{array}$ & $\begin{array}{r}
0\\
0\\
0
\end{array}$ & $\begin{array}{r}
1\\
0\\
-1
\end{array}$ & $\begin{array}{r}
0\\
0\\
0
\end{array}$ & $\begin{array}{r}
1\\
1\\
1
\end{array}$ & $\begin{array}{r}
0\\
0\\
0
\end{array}$\tabularnewline
\hline
\end{tabular}
\label{tab:table5p1}
\par\end{raggedright}
\caption{Massless fermion and bosons states in (5+1)-d extension, momentum along ${\pm\bf {\hat z}}$, with projection  given by the lepton number ${\mathcal P}_P=L$, under the operators SU(2)$_L$ $I_3$ component, hypercharge $Y$, charge $Q=I_3+\frac{1}{2}Y$, the lepton number $L$, helicity $\frac{i}{2}L\gamma^{1}\gamma^{2}$, and  chirality $L\tilde{\gamma}_{5}$ (the coordinate dependence is omitted.) }
\end{table}

\normalsize




 By identifying elements between the extended spin space and standard Lagrangian terms, Ref. \cite{Besprosub} set thumb rules to derive some gauge-invariant terms. For rigor's  sake, and to test the model's reach, it is desirable to obtain such terms within the model's algebra. Next, we translate the field information that emerges from the extended-spin space, to  derive an interactive gauge theory. First, we write fields in the  extended-spin basis; similarly, the symmetry generators are written in a standard representation; finally, invariant terms are constructed, and shown to be equivalent to field-theory Lagrangian contributions.

As derived in Section 2, and exemplified above, it is  possible to
 write fundamental fields using  as basis products of matrices conformed of  Lorentz and scalar group representations.
  Indeed,
the commuting  property of  the respective degrees of freedom  allows for states and operators to be written in the form
$ {\mathcal C}_{4}{ \bigotimes }{\mathcal S}_{N-4}$;
 explicitly, $\Psi=M_1  M_2$, where
 \begin{eqnarray}
\label {Separation}
   M_1 \in {\mathcal C}_{4} \ \  {\rm and} \ \    M_2 \in  {\mathcal S}_{N-4} .
\end{eqnarray}
An expression with elements of each set is possible through their  passage   to each side, using    commutation or  anticommutation rules.


\subsection {Fields' construction}

In the presence of interactions, free fields give way  to more general expressions of    fermion and boson fields,  keeping their transformation properties:
\begin{description}
\item [ Vector field]
\begin{eqnarray}
\label {AmuExpandfinal}
 A_\mu^a (x) \gamma_0\gamma_\mu I_a,
\end{eqnarray}
\end{description}
where $\gamma_0\gamma_\mu \in{\mathcal C}_{4}$, and  $I_a \in {\mathcal S}_{N-4}^\prime$ is a generator of a given
unitary group,   and  ${\mathcal S}_{N-4}^\prime$ is defined in Eq. \ref{Sprima}.
\begin{description}
\item[ Scalar field]
\begin{eqnarray}
\label {phiExpandfinal}
 \phi^a (x) \gamma_0  M_a^S.
\end{eqnarray}
 \end{description}
\begin{description}
\item [  Fermions]
\begin{eqnarray}
\label {Fermion}
 \psi^a_\alpha (x) L^\alpha  P_F  M_a^F ,
\end{eqnarray}
 \end{description}
where $M_a^S$,$M_a^F \in {\mathcal S}_{N-4}$ are, respectively, scalar and fermion components, and  $L^\alpha$ represents a spin component;  for example,   $L^1=(\gamma_1+ i \gamma_2) $,   $P_F$ is a projection operator of the type in Eq. \ref{Lorentgen}, such that
\begin{eqnarray}
\label {transfoyy} P_F \gamma_\mu= \gamma_\mu P_F^c,\end{eqnarray}
and we use the complement $P_F^c  = 1- P_F$, so that a Lorentz transformation with $P_F  \sigma_{\mu\nu}$, will describe fermions, as argued in Section 3.  The simplest example for an operator satisfying such conditions is  $P_F = (1-\tilde\gamma_5 )/2$ \cite{Jaime,JaimeB}, used by the fermion doublet on Table 2.
 By the argument after Eq.  \ref{Lorentgen},   the  fundamental-representation state is   derived from
  the  trivial right-hand action of the operator within the transformation rule in Eq. \ref{transfoyy}.
   The matrix  entitles spurious ket states contained in the Lorentz-scalar matrices in Eq. \ref{Separation}.


\begin{description}
\item [ Antisymmetric tensor]
\end{description}

 It is also obtained, however, but as it leads to non-renormalizable interactions, it  will hence be  omitted.

 \subsection {Symmetry Transformations}
We now describe different types of transformations that act as in Eq.  \ref{transfoGen}:
 \subsubsection* {Lorentz Transformation}

 \begin{eqnarray}
\label  {transfoLorentz}
 U=\exp (-\frac{i}{4} {\mathcal P}_P w^{\mu\nu}\sigma_{\mu\nu}),
\end{eqnarray}
where $ \sigma_{\mu\nu}$ is given in Eq. \ref{sigmamunu}, $w^{\mu\nu}$ are parameters   and  $ {\mathcal P}_P $ is the scalar projector in Eq.  \ref{Lorentgen}.
  \subsubsection*{Gauge Transformation}
  \begin{eqnarray}
\label  {transfoGauge}
 U=\exp [-i  I_a  \alpha_a(x)],
\end{eqnarray}
where $I_a\in  {\mathcal S}_{N-4}^\prime$, and  $\alpha_a(x)$ are arbitrary functions.
The  unitary-group representations $\bar N\bigotimes N$,  based on elements in the fundamental representation and its conjugate, denoted by $N$, $\bar N$, respectively, are implicit from the $| a\rangle \langle b |$ matrix construction in Eq. \ref{Jaimeqnext}; these
  include the singlet,     and the
     fundamental (expressed in $I_a$) representations,  and similarly those obtained by $  N\bigotimes N$ (see, e.g.,  \cite{bespro9m1}.)





\section {Lagrangian connection}

Historically, it is known that Maxwell's equations can be formulated in terms of a Dirac basis\cite{Bargmann}. In our case, the fields    within the extended-spin basis can be used to construct  a   standard-formulation Lagrangian.\footnote{Alternatively,  the fields'\   Lagrangian describing them  can be reinterpreted in
terms of this basis.}
   This amounts to using elements   with a well-defined  group structure to get Lorentz-scalar gauge-invariant
combinations.
 Choosing scalar elements that result from the direct product in Eq. \ref{dotprod},
  one obtains an interactive theory, as the same  particle content is maintained.  In this way, choices of Lagrangians are  constrained by the same conditions as in quantum field theory,
  as  renormalizability and quantization.

We proceed by first  constructing   matrix elements containing the vector field, together with either fermion  or bosons  fields, and then converting them to expressions in terms of states'\
 associated  bras or kets.        Under    Lorentz and gauge-group transformations  of the extended spin space, invariant elements are obtained by taking  the trace. The  latter extracts   the identity-matrix coefficient,  leading  to  the usual Lagrangian components.
The invariance under transformations in Eq. \ref{transfoGen}    can be verified independently, using the    separation   in Eq. \ref{Separation} into Lorentz and scalar symmetries;
the invariance  will be shown  for linear (vector-fermion) or bilinear (vector-scalar) objects,
 with input from
  Eqs. \ref{AmuExpandfinal}-\ref{Fermion}:

\subsection {Fermions}
A gauge-invariant fermion-vector   interaction term results, by  adding to the fermion free-term Lagrangian (that implies   the Dirac equation   \ref{JaimeqDi})  the vector-term contribution in  Eq. \ref{AmuExpandfinal}
\begin{eqnarray}
\label {Bilin}
 \frac{1}{N_f} {\rm tr}\Psi^\dagger\{  [ i\partial_\mu I_{den}+g A_\mu^a (x) I_a ] \gamma_0\gamma^\mu-   M\gamma_0  \}\Psi P_f,
\end{eqnarray}
where $\Psi$ is a  field representing  in this case spin-1/2 particles; spin-1 terms are treated below.
 $ I_a$  is the group generator in a given   representation, $g$ is the coupling constant, $N_f$ contains the normalization (and similar terms below),
and $I_{den}$ the  identity scalar group operator in the same representation (which will be omitted hence). An operator $P_f$ is introduced to avoid cancelation of non-diagonal fermion  elements. For example,
\begin{eqnarray}
\label  {Pf}
 P_f=\frac{1}{\sqrt{2}}[(1+i) (I+\gamma^0 \gamma^2)+ \gamma^5 \gamma^6+\gamma^0 \gamma^2 \gamma^5 \gamma^6]
\end{eqnarray}
 as $[P_f,L]=[P_f,(1-\tilde\gamma_5)L]=0$, provides a non-trivial combination with  the correct quantum numbers for the fermion pair
$\Psi_a  P_f\Psi^\dagger_b$ (with $\Psi_a,  \Psi_b$ either doublet or singlet fermions, on Table 2), and maintains their  normalization, spin,   lepton and electroweak representation.

 As explained after Eq.   \ref{Fermion}, Lorentz and scalar operators act non-trivially only from one side.  
 Given the action of projection operators ${\mathcal P}_{S}$ ${\mathcal P}_{P}, $ the transformation in Eq.  \ref{transfoGen} becomes
\begin{eqnarray}
\label  {transfoGenFermi}
 \Psi\rightarrow U \Psi.
\end{eqnarray}
Eq.   \ref{Bilin} is invariant under the Lorentz transformation in Eq. \ref{transfoLorentz}, provided the vector field transforms as
\begin{eqnarray}
\label {transformsLor}
A_\mu^a (x) I_a  \rightarrow \Delta_\mu^{\ \ \nu} A_\nu^a (x) I_a,
\end{eqnarray}
where we use the identity relating the spin representation of the Lorentz group in
\begin{eqnarray}
\label {LorentzIden}
U  \gamma^\mu  U^{-1}=({\Delta^{-1}})^\mu_{\ \ \nu} \gamma^\nu ,
\end{eqnarray}
and  $\Delta^\mu_ {\ \ \nu} $ is a $4\times 4$ Lorentz transformation matrix transforming a coordinate
as $x^\mu   \rightarrow \Delta^\mu_{\ \ \nu} x^\nu$.
The equation is also invariant under local transformation in Eq. \ref{transfoGauge}, under the condition the vector field transforms as
\begin{eqnarray}
\label {transforms}
A_\mu^a (x) I_a  \rightarrow U A_\mu^a (x) I_a U^\dagger -\frac{i}{g}(\partial_\mu U) U^\dagger ,
\end{eqnarray}

The trace  in Eq. \ref{Bilin} can be expressed in terms of states,  as we rely on the expansion in Eq. \ref {Jaimeqnext} for fields $\Psi$.
The   fermion field in Eq. \ref{Fermion},  with matrix elements $\gamma\delta$, is expressed as
$ [ \psi^a_\alpha (x) L^\alpha  P_F  M_a^F] _{\gamma\delta}=\sum_{\eta} (L^\alpha P_F)_{\gamma\eta}  (M_a^F)_{\eta\delta} \langle\alpha a x |\Psi\rangle  $, where
  $(L^\alpha P_F)_{\gamma\eta}=\langle\gamma |\alpha\rangle  \sum_\beta   d_{\alpha  \beta} \langle  \beta |\eta\rangle$,
  $(M_a^F)_{\eta\delta}=\langle\eta |a\rangle  \sum_b  f_{ab} \langle b |\delta\rangle$, with  $d_{\alpha \beta}, f_{ab} $ c-numbers,  and  where the fermion field $  \psi^a_\alpha (x)   =\langle\alpha a x |\Psi\rangle $ takes account of spin and  scalar degrees of freedom.
 We choose commuting scalar and spin matrices  as
basis elements, as the fermion singlet on Table 2;     we use the separability
property\cite{CohenTanu} of the generators  (for anticommuting matrices, as for the doublets,
 each bilinear term is separated),
 as the normalization condition, in Eq. \ref{unitarity}, cancels the ket-bra:  for the spin component,
$ L_\alpha P_F \tilde L_f ( L_\beta P_F)^\dagger  = {( L_\alpha P_F)}_{\gamma\delta}(\tilde L_f)_{\delta\gamma}{(L_\beta P_F)} ^*_ {\epsilon\gamma}=\langle\gamma |\alpha\rangle\langle\beta |\epsilon\rangle $, and similar calculation involving the electroweak states (here $\tilde L_f$ is a reduced operator $  L_f$  acting on spin degrees of freedom.)
This implies Eq. \ref{Bilin} can be written
\begin{eqnarray}
\label {BilinNoTrace}
{ \psi^b_\beta}^ \dagger(x)    \{   [ i\partial_\mu I_{bc}+g A_\mu^a (x) (I_a)_{bc} ](  \gamma_0\gamma^\mu )_{\beta\alpha}  - M I_{bc}(\gamma_0)_{\beta\alpha}  \}  \psi^c_\alpha (x).
\end{eqnarray}
Thus, Table 2 leads to the fermion electroweak SM Lagrangian contribution\cite{Glashow}-\cite{GlashowC}, also derived heuristically in Refs.   \cite{Besprosub} and \cite{BesproMosh} .
\begin{eqnarray}
\label {BilinNoTracElectroweak}
 {\bar {\boldmathPsi_l} }       [ i\partial_\mu + \frac{1}{2}g\tau^a  W_\mu^a(x)  - \frac{1}{2}g'  B_\mu (x) ]   \gamma^\mu    {\boldmathPsi_l}+{ \bar {\psi_r} }      [ i\partial_\mu   - g'  B_\mu (x) ]    \gamma^\mu   { \psi_r},
\end{eqnarray}
which contains a left-handed  hypercharge $Y_l=-1$ SU(2) doublet $\boldmathPsi_l$, and right-handed  $Y_r=-2$ singlet $ \psi_r$, and the corresponding gauge-group vector bosons and coupling constants are, respectively, $B_\mu (x)$, $W_\mu^a(x)$, and  $g$, $g'$.

\subsection{Spin-0 Boson}
 A Lorentz-invariant  interaction term between  vector and scalar  fields is constructed by applying twice the operator contained within the state $\Psi$ in Eq. \ref{Bilin},   removing the $\gamma_0$ matrix, following the Klein Gordon equation:
\begin{eqnarray}
\label {BilinNoTraceTR}
{\rm tr}\ \frac{1}{N_B}\Psi^\dagger  [  i  \partial_\nu I_{den}+g A_\nu^b (x) I_b ]    \gamma^\nu  \gamma^\mu [ i\partial_\mu I_{den}+g A_\mu^a (x) I_a ]   \Psi ,
\end{eqnarray}
where the transformation in Eq. \ref{transfoGen} is now used in the guise
 $\Psi\rightarrow U \Psi U^{-1}$, and the 4-d $\gamma_\mu$ are positioned in near pairs to maintain the generators $I_a$ relations (see also the vector term in Eq. \ref{vectortraceFmunu};)
this expression  applies to the Lorentz transformation as in Eq. \ref{transfoLorentz}.
The final expression is obtained by applying the equality $\gamma_\mu \gamma_\nu =g_{\mu\nu}-i \sigma_{\mu\nu}$, as the only symmetric term  $[ i\partial _\nu I_{den}+A_\nu^b (x) I_b ] [ i\partial_\mu I_{den}+A_\mu^a (x) I_a ] =\frac{1}{2} \{   i  \partial _\nu I_{den}+A_\nu^b (x) I_b  ,   i\partial_\mu I_{den}+A_\mu^a (x) I_a  \}$ survives the renormalizability demand.\footnote{Where we use the operator equality $AB=\frac{1}{2} [A,B]+\frac{1}{2} \{A,B\}$, and the antisymmetric term cancels through the trace on the (3+1)-d spinor indices} A similar expansion as for the fermion field in Eq.  \ref{Bilin} can be performed, the  two $\gamma_0$ matrices    in the field terms in Eq. \ref{phiExpandfinal},  contained in  $\Psi\Psi^\dagger$,  lead to the identity matrix within  the trace. The vector mass term resulting from the Higgs mechanism was related to mass operators within the spin-extended space, and used to connect it to the SM in Ref. \cite{Besprosub}.

\subsection{Vector Boson}

We use invariant components for the vector field contained in Eq. \ref{AmuExpandfinal} to construct its kinetic-energy term, and
  we extract the antisymmetric part
\begin{eqnarray}
\label {vectortraceFmunu}
[ i\partial_\nu I_{den}+g A_\nu^b (x) I_b ]    [ i\partial_\mu I_{den}+g A_\mu^a (x) I_a ] \frac{i}{2} [\gamma^\nu ,\gamma^\mu ]=F_{\mu\nu}^a I_a \frac{i}{2} [\gamma^\nu ,\gamma^\mu ],
\end{eqnarray}
where by taking the antisymmetric tensor  $[\gamma^\nu ,\gamma^\mu ]$ we extract
$F_{\mu\nu}^a=\partial_\mu A^a_\nu-\partial_\nu A^a_\mu+g c^{a b d}A^b_\nu A^d_\mu $,
and $ c^{a b d}$ are the structure constants of the group $[I_b ,I_d]  =i c^{a b d} I_a$,

We show a particular term that reproduces the kinetic vector contribution, which eliminates non-renormalizable higher-derivative terms.  A scalar contribution is constructed from the  contraction of the two terms
 \begin{eqnarray}
\label {vectortraceFmunuContract} \frac{1}{N_A}{\rm tr }   F_{\mu\nu}^a I_a \frac{i}{2} [\gamma^\nu ,\gamma^\mu ] F_{\rho\sigma}^b I_b \frac{i}{2} [\gamma^\rho ,\gamma^\sigma ].  \end{eqnarray}
 From the 4-d  trace relation
  \begin{eqnarray}
 \label {traFMunu}
{ \rm tr }\gamma^\mu\gamma^\nu\gamma^\rho\gamma^\sigma=g ^{\mu \nu}g^ {\rho \sigma}-g ^{\mu\rho }g^ {\nu \sigma}+g ^{\mu \sigma}g^ {\nu \rho}
 \end{eqnarray}
the trace is reduced to the  anti-symmetrized combination,
$-g ^{\mu\rho }g^ {\nu \sigma}+g ^{\mu \sigma}g^ {\nu \rho}$.
We finally get the known expression for the kinetic term $-\frac{1}{4}F_{\mu\nu}^a F^{\mu\nu\ a}  $.
The expression in Eq. \ref{vectortraceFmunuContract}  may be also derived  from the original corresponding standard Lagrangian\cite{Besprosub}.

\section {(7+1)-dimensional electroweak spinors:  mass term and hierarchy  }

We  described fermion-vector and vector-scalar Lagrangian contributions,  and in this Section we deal with fermion-scalar terms.  An inherent aspect of the $(5+1)$-d space   is the impossibility of defining  fermion masses for both flavor-doublet components.   The $(7+1)$-d  space  allows  for charge $2/3$ and $-1/3$  terms, associated to   quarks,  and  charge $-1$    and neutral leptons.  We concentrate on quarks, while the results of this section can be equally applied to leptons.

The
baryon-number operator
$B=\frac{1}{6}(1-i\gamma_{5}\gamma_{6})$ conforms a  spin-space partition  obtained with   the  additional Clifford members $\gamma_{7}$, $\gamma_{8}$.
It  allows  for  quark symmetry  generators
that include   the hypercharge
$Y=\frac{1}{6}\left(1-i\gamma_{5}\gamma_{6}\right)[1+i\frac{3}{2}(1+\tilde{\gamma}_{5})\gamma_{7}\gamma_{8}]$,
 the weak SU(2)$_L$ terms
   \begin{eqnarray} \label{isospinquarks}
\begin{array}{c}
{\displaystyle {\displaystyle I_{1}=\frac{i}{8}(1-\tilde{\gamma}_{5})(1-i\gamma_{5}\gamma_{6})\gamma^{7}},}\\
\\
{\displaystyle I_{2}=\frac{i}{8}(1-\tilde{\gamma}_{5})(1-i\gamma_{5}\gamma_{6})\gamma^{8}},\\
\\
{\displaystyle I_{3}=\frac{i}{8}(1-\tilde{\gamma}_{5})(1-i\gamma_{5}\gamma_{6})\gamma_{7}\gamma_{8}},
\end{array}
  \end{eqnarray} flavor generators,
and the Lorentz generators, with   spin component    $3B\sigma_{\mu\nu}$, projected by $B$, using  Eq. \ref{sigmamunu}. As required, $[Y,I_i]=[B,Y]=[B,I_i]=0$, and all quarks are associated the baryon number 1/3 ($-1/3$ for  antiparticles.)

Examples of quark massless basis states,  expressed as in Eq. \ref{Fermion},  are summarized on Table  3, for both u and d-type quarks, with their quantum numbers.
The spin component along $\hat {\bf z}$,
$
i\frac{3}{2}B\gamma^{1}\gamma^{2},
$ is used.
 Only one polarization and two flavors are shown, as a
more thorough treatment of the fermion flavor states  will be given elsewhere\cite{Romero}.
\begin{table}[ht]
\begin{centering}
\begin{tabular}{|c|c|c|c|}
\hline
hypercharge  $1/3$   left-handed doublet & $I_{3}$ & $Q$ & $\frac{3i}{2}B\gamma^{1}\gamma^{2}$\tabularnewline
\hline
\hline
$ {     \bolditQ}_L^1=\left(\begin{array}{lcr}
   U_{L}^1 \\
  D_{L}^1\\
 \end{array}\right)=\left(\begin{array}{c}
 \frac{1}{16}\left(1-\tilde{\gamma}_{5}\right)\left(\gamma^{5}-i\gamma^{6}\right)\left(\gamma^{7}+i\gamma^{8}\right)\left(\gamma^{0}+\gamma^{3}\right)\\
\frac{1}{16}\left(1-\tilde{\gamma}_{5}\right)\left(\gamma^{5}-i\gamma^{6}\right)\left(1-i\gamma^{7}\gamma^{8}\right)\left(\gamma^{0}+\gamma^{3}\right)
\end{array}\right)$ & $\begin{array}{r}
1/2\\
-1/2
\end{array}$ & $\begin{array}{r}
2/3\\
-1/3
\end{array}$ & $\begin{array}{r}
 1/2\\
 1/2
\end{array}$\tabularnewline
\hline
\end{tabular}
\par\end{centering}
\center{(a)}
\begin{centering}
\begin{tabular}{|c|c|c|c|}
\hline
  $I_{3}=0$   right-handed singlet & $Y$ & $Q$ & $\frac{3i}{2}B\gamma^{1}\gamma^{2}$\tabularnewline
\hline
\hline
$\begin{array}{c}
U_R^1=\frac{1}{16}\left(1+\tilde{\gamma}_{5}\right)\left(\gamma^{5}-i\gamma^{6}\right)\left(\gamma^{7}+i\gamma^{8}\right)\gamma^{0}\left(\gamma^{0}+\gamma^{3}\right)\\
D_R^1 =\frac{1}{16}\left(1+\tilde{\gamma}_{5}\right)\left(\gamma^{5}-i\gamma^{6}\right)\left(1-i\gamma^{7}\gamma^{8}\right)\gamma^{0}\left(\gamma^{0}+\gamma^{3}\right)
\end{array}$ & $\begin{array}{r}
4/3\\
-2/3
\end{array}$ & $\begin{array}{r}
2/3\\
-1/3
\end{array}$ & $\begin{array}{r}
 1/2\\
 1/2
\end{array}$\tabularnewline
\hline
\end{tabular}
\end{centering}
\center {(b)}
 \caption {(a)Massless  left-handed quark weak isospin doublet, and (b) right-handed  singlets, with momentum along $\pm{\bf {\hat z}}$.}
\label{tab:tablequarks}
\end{table}
\baselineskip 22pt\vfil\eject \noindent

On  Table 4 we also present two scalar elements as  in Eq.  \ref{phiExpandfinal}, whose quantum numbers associate them to the Higgs doublet.  These are unique within the  (7+1)-d space\cite{Romero}.
\begin{table}[ht]
\begin{centering}
\begin{tabular}{|c|c|c|c|c|}
\hline
$0$ baryon-number  scalar    & $I_{3}$ & $Y$ & $Q$& $\frac{3 i}{2}B\gamma^{1}\gamma^{2}$\tabularnewline
\hline
\hline
${ \boldmathphi}_1$=$\left(\begin{array}{c}
\phi_{1}^{+}\\
\phi_{1}^{0}
\end{array}\right)$ =
$\left(\begin{array}{c}
 \frac{1}{8}\left(1-i\gamma_{5}\gamma_{6}\right)\left(\gamma^{7}+i\gamma^{8}\right)\gamma^{0}\\
 \frac{1}{8}\left(1-i\gamma_{5}\gamma_{6}\right)\left(1+i\gamma_{7}\gamma_{8}\tilde{\gamma}_{5}\right)\gamma^{0}
\end{array}\right)$ & $\begin{array}{r}
 1/2\\
 -1/2
\end{array}$ & $1$ & $\begin{array}{r}
1\\
0
\end{array}$& $0$\tabularnewline
\hline
${ \boldmathphi}_2$=$\left(\begin{array}{c}
 \phi_{2}^{+} \\
\phi_{2}^{0}
\end{array}\right)$ =
$\left(\begin{array}{c}
 \frac{1}{8}\left(1-i\gamma_{5}\gamma_{6}\right)\left(\gamma^{7}+i\gamma^{8}\right)\tilde{\gamma}_{5}\gamma^{0}\\
 \frac{i}{8}\left(1-i\gamma_{5}\gamma_{6}\right)\left(1+i\gamma_{7}\gamma_{8}\tilde{\gamma}_{5}\right)\gamma^{7}\gamma^{8}\gamma^{0}
\end{array}\right)$ & $\begin{array}{r}
1/2\\
-1/2
\end{array}$ & $1$ & $\begin{array}{r}
1\\
0
\end{array}$& $0$\tabularnewline
\hline
\end{tabular}
\label{tab:tableHiggs}
\par\end{centering}
\caption{Scalar Higgs-like  pairs}
\end{table}
\normalsize
The combination $a{ \boldmathphi}_1+b{ \boldmathphi}_2$ for arbitrary real $a$, $b$ is classified with the chiral projection operators $L_5=\frac{1}{2}(1-\tilde \gamma_5)$,  $R_5=\frac{1}{2}(1+\tilde \gamma_5)$, giving $R_5({ \boldmathphi}_1+{ \boldmathphi}_2)L_5={ \boldmathphi}_1+{ \boldmathphi}_2$, $L_5({ \boldmathphi}_1+ { \boldmathphi}_2)R_5=0$, $L_5({ \boldmathphi}_1- { \boldmathphi}_2)R_5={ \boldmathphi}_1-{ \boldmathphi}_2$,
$R_5({ \boldmathphi}_1- { \boldmathphi}_2)L_5=0$. This leads to the
 gauge-invariant  Lagrangian
\begin{eqnarray}
\label {BilinHiggs}
 \frac{1}{N_f} {\rm tr}\{ [ m_U  { \Psi_R^U}^\dagger(x) [  { \boldmathphi}_1(x) +{ \boldmathphi}_ 2 (x)]{ \boldmathPsi}_L^Q(x) +m_D {{ \boldmathPsi}_L^Q}^\dagger(x) [{ \boldmathphi}_1 (x)-{ \boldmathphi}_ 2 (x)]  \Psi_R^D (x)] P_f \}+\{ cc \},
\end{eqnarray}
in terms of the scalar fields
${ \boldmathphi}_1(x)$=$\left(\begin{array}{c}
 \psi_1^{+}(x)\phi_{1}^{+} \\
 \psi_1^{0}(x)\phi_{1}^{0}
\end{array}\right)$ and ${ \boldmathphi}_2(x)$=$\left(\begin{array}{c}
  \psi_2^{+}(x)\phi_{2}^{+}\\
 \psi_2^{0}(x)\phi_{2}^{0}
\end{array}\right)$, and
  quark fields
 ${ \Psi_R^U}(x)=\sum_\alpha \psi_{UR}^\alpha(x)U^\alpha_R$, \ ${ \Psi_R^D}(x)=\sum_\alpha \psi_{DR}^\alpha (x)D^\alpha_R$, \   and
   $\boldmathPsi_L^Q(x)=\sum_\alpha \left(\begin{array}{lcr}
   \psi_{UL}^\alpha(x) U_{L}^\alpha \\
 \psi_{DL}^\alpha(x) D_{L}^\alpha \\
  \end{array}\right) $,
where $P_f$ is a projection operator,$\alpha$ is a spin component, and, with hindsight, we assign the masses
 \begin{eqnarray}
\label {massinter} m_U=(a+b)/2, \    \    m_D=(a-b)/2 .
 \end{eqnarray}
 Eq. \ref{BilinHiggs}'s  configuration makes manifest  the required gauge symmetries:  SU(2)$_L$ for a   field $\Psi(x)$
  \begin{eqnarray}
\label {genSU2YSymmetry}
  \Psi(x)\rightarrow e^{i \sum_c\alpha_c(x) I_c} \Psi(x)  e^{-i \sum_d\alpha_d(x) I_d}
\end{eqnarray}
leads to the non-trivial transformations
\begin{eqnarray}
\label {SU2YSymmetry}
  \boldmathphi_1(x)-\boldmathphi_2(x)\rightarrow  e^{i \sum_c\alpha_c(x) I_c} [ \boldmathphi_1(x)-\boldmathphi_2 (x)] \\
  \boldmathphi_1(x)+\boldmathphi_2(x)\rightarrow   [ \boldmathphi_1(x)+\boldmathphi_2(x)] e^{-i \sum_d\alpha_d(x) I_d} \\
  \boldmathPsi_L^Q(x) \rightarrow e^{i \sum_c \alpha_c(x) I_c}  \boldmathPsi_L^Q(x),
\end{eqnarray} and for the   U(1)$_Y$ transformation
\begin{eqnarray}
\label {generalU1}
 \Psi(x)\rightarrow e^{i \alpha_Y(x) Y}  \Psi(x) e^{-i \alpha_Y(x) Y}
\end{eqnarray} implies
\begin{eqnarray}
\label { U1YSymmetry}
  \boldmathphi_1(x)-\boldmathphi_2(x)\rightarrow   e^{i  \alpha_Y(x)  1/3 } [ \boldmathphi_1(x)-\boldmathphi_2(x)]e^{i  \alpha_Y(x) 2/3 } \\
   \boldmathphi_1(x)+\boldmathphi_2(x)\rightarrow   e^{ i  \alpha_Y  4/3 } [ \boldmathphi_1(x)+\boldmathphi_2(x)]e^{-i  \alpha_Y(x) 1/3 } \\
  \boldmathPsi_L^Q(x) \rightarrow   e^{i \alpha_Y(x)  1/3  } \boldmathPsi_L^Q(x) \\
  \Psi_R^U(x) \rightarrow   e^{i \alpha_Y(x) 4/3  } \Psi_R^U(x) \\
   \Psi_R^D(x) \rightarrow    e^{ -i \alpha_Y(x) 2/3  } \Psi_R^D(x).
\end{eqnarray}
These relations imply scalar components are connected to the SM Higgs ${\bf H}$  through the assignments
\begin{eqnarray}
\label {correspH}
 { \bf H(x)} \sim \boldmathphi_1(x)-\boldmathphi_2(x)\\ \label {correspHo}
 \tilde{\bf H}^\dagger(x) \sim \boldmathphi_1(x)+\boldmathphi_2(x) ,
\end{eqnarray}
where the conjugate representation  corresponds to $\tilde{\bf H}(x)=i I_2 {\bf H}^*(x)$,   a unitary transformation connects them
to their conjugates, e.  g. (see Table 4),
\begin{eqnarray}
\label {transforH}
{\phi_{1}^{+}}^\dagger+{\phi_{2}^{+}}^\dagger=-2 I_2\gamma_2(\phi_{1}^{0}+\phi_{2}^{0})^*\gamma_2,
\end{eqnarray}
 and the Dirac representation for the   $\gamma_\mu$ matrices fixes   charge conjugation.

After the
Higgs mechanism\cite{higgsMech}-\cite{higgsMechC}, only neutral fields survive, and    the  same basis as Table 4 for the  vacuum expectation value is used, leading
 to the  mass Lagrangian
 \begin{eqnarray}
\label  {HiggsComponents}
  H_v= a \phi_1^0 + b  \phi_2^0   + a  {\phi_1^0}^\dag + b   {\phi_2^0}^\dagger ,
  \end{eqnarray}
This term produces fermion eigenstates and masses from the   Yukawa coupling parameters
through the relations
\begin{eqnarray}
\nonumber
 H_v U_{M}^1=m_U U_{M}^1, \ \   H_v  U_{M}^{c1} =-m_U U_{M}^{c1}  , \\
   \label{MassiveEigenvalues}  H_v D_{M}^1 =m_D D_{m}^1 , \  \  H_v D_{M}^{c1} =-m_D D_{M}^{c1},
 \end{eqnarray}
 where $U_{M}^{c1}$, $ D_{M}^{c1}$ correspond to negative-energy solution states (and similarly   for opposite spin components.)
These   states  are listed on Table  5 with their quantum numbers;\footnote{The fermion states shown can be interpreted as either massive quarks or massive leptons (charged particle and neutrino pairs), according to the choice of the $Y$ operator.}
only two flavors are shown.
\begin{table}[ht]
\noindent \begin{centering}
\begin{tabular}{|c|c|c|c|}
\hline
  massive quarks & $H_{v}$ & $Q$ & $\frac{3i}{2}B\gamma^{1}\gamma^{2}$\tabularnewline
\hline
\hline
$ U_{M}^1 =\frac{1}{\sqrt{2}} ({U_L^1}+{U_R^1} )$ & $m_U$ & $2/3$ & $ 1/2$\tabularnewline
\hline
$D_{M}^1=\frac{1}{\sqrt{2}} ({D_L^1}-{D_R^1} )$ & $m_D$ & $-1/3$ & $ 1/2$\tabularnewline
\hline
$ U_{M}^{c1} =\frac{1}{\sqrt{2}} (U_{L}^1-U_{R}^1 )$ & $-m_U$ & $2/3$ & $ 1/2$\tabularnewline
\hline
$D_{M}^{c1}=\frac{1}{\sqrt{2}}({D_L^1}+{D_R^1})$ & $-m_D$ & $-1/3$ & $ 1/2$\tabularnewline
\hline
\end{tabular}
\par\end{centering}
\caption{Massive quarks eigenstates of $H_v$}
\end{table}
\normalsize
 The role played by $m_U$, $m_D$ in Eq. \ref{MassiveEigenvalues} confirms their mass  interpretation  in Eq. \ref{massinter}.   In addition, the  particular dependence on the $a$, $b$ parameters  implies
 a  flavor-doublet mass  hierarchy effect, if they represent a comparable large scale O($a ) \simeq$O( $ b$).
 This interpretation is supported  by the connections among the Higgs components   on Table 4
  $ {\boldmathphi }_2= \gamma_5 {\boldmathphi  }_1$,   and as ${\boldmathphi }_2 $ can be generated from ${\boldmathphi }_1$ by the  transformation
 \begin{eqnarray}
 \label  {UnitaryGamma5}
 {\boldmathphi  }_2=-i e^{i\beta \gamma_5} {\boldmathphi  }_1  e^{-i\beta \gamma_5}
 \end{eqnarray}
for $\beta=\pi/4$;
 further on,  by a compositeness property,  as one may construct the Higgs wave function from the fermions. This is shown in the relations
 \begin{eqnarray}
\label  {HiggsComponentsU}
{\phi_1^0}^\dagger+{\phi_2^0}^\dagger=U_{L}^1 {U_{R}^1}^\dagger+U_{L}^2 {U_{R}^2}^\dagger
\end{eqnarray}
 \begin{eqnarray}
\label  {HiggsComponentsD}
{\phi_1^0}^\dagger-{\phi_2^0}^\dagger=-D_{L }^1 {D_{R }^1}^\dagger-D_{L }^2 {D_{R }^2}^\dagger ,
\end{eqnarray}
and the second spin component may be obtained by flipping the spin, for example,
 $D_{L }^2 =\frac{3 i }{2} B (\gamma_2 \gamma_3-i \gamma_3 \gamma_1) D_{L }^1$.










\section{Conclusions}
This paper presented two related themes: one formal, dealing with translating a  previously proposed SM extension to a Lagrangian formalism, and the other phenomenological, dealing with deriving a hierarchy effect from the model.
It explained steps  aimed at the model's formalization,  providing  a field-theory formulation;  the final objective is to use its restrictions to obtain SM information. Conversely, a field theory can be formulated in this basis, which may provide insight into the symmetries and representations used.

A matrix space is used in which  both   symmetry generators and fields are formulated.
For given dimension, a chosen non-trivial  projection operator ${\mathcal P}_P$   constrains the  matrix space,     determining  the   symmetry groups, and the  arrangement  of fermion  and boson  representations.
 In particular,  spin-1/2, and 0 states   are  obtained in the fundamental representation of  scalar groups and spin-1 states in the adjoint representation.
 After expressing fields within this basis,
  a gauge-invariant field theory is constructed,   based on the Lorentz and obtained scalar symmetries.

  Features obtained from the  (5+1)-d extension are formulated through a Lagrangian:
  the   gauge symmetry
$\rm SU(2)_L \bigotimes U(1)_Y $  and global lepton $\rm U_{Le}(1)$   groups
with the   vector
bosons  associated   to  SU(2)$_L$, acting only on the model's predicted representations: left-handed fermions;
  a scalar doublet    associated to  a Higgs particle; leading to
  scalar vector and fermion vertices. Special features emerge in the Lagrangian construction, as the need of a projection operator and Dirac-matrix rules to maintain Lorentz invariance.
  Within the (7+1)-d case, we showed a pair of Higgs-like scalars    induce hierarchy in the masses of   flavor-doublet fermions, confirming the model's predictive power.

The paper's SM extension satisfies basic requirement of correct symmetries, including Lorentz and gauge ones, description of SM particles, and field-theory formulation, in addition to its SM prediction provision (the latter two is what the paper deals with.) This supports the view that it is an extension worth considering.

With  the  Poincar\'e  and SM-gauge symmetric  Lagrangian
presentation of the model, renormalization and quantization conditions can be
applied, leading to a quantum field theory formulation.

A future goal is to apply this   framework to supersymmetry.  The latter has in common with the extended spin representations classified by   a Clifford algebra with Lorentz indexing. This   suggests a closer connection between these frameworks.  As restrained matrix spaces provide information on   fundamental interactions  and physical-particle representations,
it is worth investigating  whether this information can be obtained within supersymmetry, with the ultimate goal of explaining the origin of interactions.

\setcounter{equation}{0}


{\bf Acknowledgements}
 The
authors acknowledge support from  DGAPA-UNAM,  project IN115111.
















%





























\begin{thebibliography}{99}



\bibitem {unification}  H.  Georgi  and S. L.  Glashow,  Phys. Rev. Lett. {\bf 32}, 438 (1974).


\bibitem {Jaime}  J.  Besprosvany,
Int. J. Theor. Phys. { \bf 39},  2797 (2000).

\bibitem {JaimeB} J. Besprosvany, Nuc. Phys. B (Proc. Suppl.) {\bf 101}, 323 (2001).

\bibitem{Coleman} S.  Coleman and J. Mandula, Phys. Rev. {\bf 159}, 1251 (1967).


\bibitem{bespro9m1}
 J. Besprosvany,
Phys. Lett. B {\bf 578} 181 (2004).



\bibitem {wess} J. Wess,  and  J.  Bagger,  {\it Supersymmetry and Supergravity}, (Princeton University Press, Princeton, 1992).


\bibitem{bespro9m1}
 J. Besprosvany,
Phys. Lett. B {\bf 578} 181 (2004).



\bibitem{Besprosub}    J. Besprosvany,  Int. J. Mod. Phys. A {\bf 20}  77 (2005).


\bibitem{BesproMosh} J. Besprosvany and R. Romero, {\it AIP  Conf. Proc.} {\bf 1323}  16 (American Institute of Physics, Melville, New York, 2010).




\bibitem {spin} K. Shima, Phys. Lett. B  {\bf 501}  237 (2001).

\bibitem {Chisholm} J. R. S.  Chisholm and R. S.  Farwell,  J. Phys A: Math. Gen.
{\bf 22} 1059 (1989).


 \bibitem{Bracic:2005ic}    A. Borstnik Bracic and
   N. S. Mankoc Borstnik,
Phys. Rev. D {\bf 74}  073013 (2006).
\bibitem {Bregar2008} G. Bregar, M. Breskvar, D. Lukman, N. Mankoc Borstnik,
 New J.Phys.
{\bf 10 }  093002 (2008).


 \bibitem {Trayling:2001kd} G. Trayling and W. B. Baylis, J. Phys A: Math. Gen.
{\bf 34} 3309 (2001). 

\bibitem {Lu2011} W. Lu,
Adv. Appl. Clifford Algebras {\bf 21}  145 (2011).

\bibitem {Nesti2009}F. Nesti,
The European Physical
Journal C {\bf 59} (3)  723 (2009).

\bibitem {Pavsic2010} M. Pavsic,
 Advances
in Applied Clfford Algebras {\bf 20 }(3-4)  781 (2010).

\bibitem {Witten} E. Witten, Nucl. Phys. B   {\bf 186}  412 (1981).









\bibitem{Bargmann} V.  Bargmann and E. P. Wigner,
Proc. Nat. Acad. Sci. (USA)  {\bf 34},  211 (1948).







\bibitem {CohenTanu} C.  Cohen-Tannoudji, B. Diu, F. Laloe,
{\it Quantum Mechanics} Vol I,
  New York, Wiley; Paris, Hermann, (1977).




\bibitem {Glashow} S. L. Glashow,  Nucl. Phys. {\bf 22},   { 579}
(1961).

\bibitem {GlashowB} S. Weinberg,  Phys. Rev. Lett.  {\bf 19}, 1264 (1967).

 \bibitem {GlashowC} A.  Salam,  in W. Svartholm (Ed.), {\it Elementary Particle
Theory}, (Almquist and Wiskell, Stockholm, 1968).



\bibitem {Romero} R. Romero and J. Besprosvany,   Electroweak quark model in extended spin space, in preparation.

\bibitem {higgsMech} F. Englert and R. Brout, Phys. Rev. Lett.  {\bf 13} 321 (1964).

\bibitem {higgsMechB} P. W. Higgs, Phys.  Lett. {\bf 12} 132 (1964).

\bibitem {higgsMechC} H. E. Haber, G. L. Kane, T. Sterling, Nucl. Phys. B {\bf 161} 493 (1979).












\end{thebibliography}
\end{document}